\documentclass{knawproc}
\usepackage{epsfig}
\def\gtwid{\mathrel{\raise.3ex\hbox{$>$\kern-.75em\lower1ex\hbox{$\sim$}}}}
\begin{document} 
\begin{opening}
\title{High Redshift Radio Galaxies at Low Redshift, and Some Other Issues}
\author{Robert Antonucci}
\addresses{Department of Physics, University of California, Santa Barbara,
California  93106}
\end{opening}

\section  {Cygnus A}

  Cygnus A is the only high redshift radio galaxy at low redshift, that is it's the only
nearby object with radio power in the range of the high redshift 3C objects.  It is clear
now that this is somewhat misleading in that Cyg A is an overachiever in the radio, and
that its actual bolometric luminosity is much more modest than this would indicate.
(This point has been explored and generalized in Barthel and Arnaud 1996; 
also see Carilli and Barthel 1996 for a detailed review of Cyg A).  But the energy
content of the lobes is famously large.  

  There is a whole history of attempts to show that Cygnus A fits the Unified Model, and
our particular contribution was detecting an apparent broad MgII line with the HST
(Antonucci, Kinney and Hurt 1994, which includes references to previous work).  The spectral
SNR was less that amazing;  furthermore an unflagged dead diode took out $\sim
12\AA$ from the
line profile; and there was an uncertain ``noise" contribution from confusing narrow lines
(gory details in Antonucci 1994).  One of the referees of our paper - the
favorable one - stated that ``only a mother could love that line."  Thus we reobserved it
with somewhat better SNR and with the bad diode flagged, and the old and new data are 
presented to the same scale in Fig 1.  Most of the bins are within the combined
1 $\sigma$
statistical
errors, and the many statistically significant wiggles are almost all present in NGC1068
as well (Antonucci, Hurt and Miller 1994).  The point is that the errors are believable,
and that the continuum should be set {\it low}.  I believe the MgII line is there and is
broader than we thought originally.  (A detailed discussion of the spectrum is in prep.)

\begin{figure}[h]
\vspace {3cm}
\caption{}
\end{figure}
  In the 1994 paper we also stated that the polarization in the UV (F320W FOC filter)
is $\sim 6 \%$, and perpendicular to the radio axis, indicating that there is a fairly large
contribution from scattered light from a quasar in this region.  This is consistent 
with the scenario of Jackson and Tadhunter 1993, among others.  Using the mighty Keck
it has finally become possible to show the broad H alpha line in polarized flux,
and it is extremely broad ($\sim 26,000$ km/sec! - Ogle et al 1997).  Ogle et al compared
the total broad H alpha and MgII fluxes in the SE component, corrected for Galactic 
reddening, and concluded that dust scattering must be important. 
(Specifically it would have to produce most of the
broad MgII.)  This was also our picture in the 1994 paper (and that of other workers).
Caveats include aperture effects and velocity ranges for integration of the line fluxes,
but the conclusion is likely to stand.

\section {Statistics of the 3CR, Unification, Possible Nonthermal AGN, and the
Role of the Mid-IR}

  Part of the Unified Model is the proposition that quasars and broad line radio galaxies
differ from narrow line radio galaxies only in orientation of the dusty tori
with respect
to the line of sight.  As explained by Barthel (1989) this idea delightfully solves
the statistical paradoxes that arise in explaining radio data for radio quasars and blazars 
alone.  The space densities and projected linear sizes of 3CR
objects in the redshift range $0.5<Z<1.0$ fit well with a simple model in which those with
torus axes within roughly 45 degrees of the line of sight correspond to the
quasars, and
the others to the (narrow line) radio galaxies.  The same is true for 3CR objects
at $Z>1$ (e.g. Singal 1993).\footnote{Pat McCarthy presented data on the large Molonglo sample, 
which has lower
luminosity than the 3CR, albeit by a modest factor.  Their data show large spreads
in size, so he argues that they neither expect nor detect a difference in projected
linear sizes for radio galaxies vs quasars.  I'd like to think (for simplicity)
that the lower luminosity of the Molonglo sources is key, and Pat acknowledges that
at the very high luminosity end there is some difference in the sizes in the expected
sense, though it may not be significant.  He wondered whether the significance of the
offsets is really high in the $Z>0.5$ objects, noting that differences in
cumulative distributions can look more significant than they are.  It's a good
lesson, and my K-S test by ruler indicates that the difference in Singal's
histogram is good at the 97.5\% confidence level for $Z>1$, but is not
significant for $0.5<Z<1.0$}

  But as Singal and other have pointed out 
(see his Figs 1 and 2),
the statistics go disastrously awry for the $Z<0.5$ objects.  There are too many radio 
galaxies per quasar, and the projected linear size difference vanishes in the median cases.

\begin{figure}[h]
\vspace {3cm}
\caption{}
\end {figure}
  What is going on with the low redshift objects?
In many cases bipolar reflection nebulae are seen around low redshift radio galaxies
(more HST UV polarization images in press by Hurt et al 1998).  There is also a growing list
with demonstrated Type 1 spectra in polarized flux; this partially overlaps the list
with the reflection nebulae.  So there is no doubt that this aspect of the Unified Model
is correct for some of them.  The simplest explanation of what may be going
wrong with the samples as a whole is
that there is an additional population of radio galaxies at these lower redshifts and
luminosities which does not participate, i.e. does not contain hidden quasars.  It has
been speculated that these are the ``optically dull" ones, without visible luminous 
high ionization narrow
line regions (e.g. Laing 1994; Barthel 1994).  
I'm sure someone has plotted the narrow line
luminosity vs radio size:  if you are out there please tell me the result.

  The low $Z$ statistics have often been presented as a counterargument to generality of 
the quasar/radio galaxy unification.\footnote{Another apparent argument for
some nonparticipating low-$Z$ radio galaxies is the weaker average IRAS far IR
fluxes found by Heckman et al 1992, 1994; however, acting on a hunch of Charlie 
Lawrence's, we have confirmed with mm data
that many of the quasars in those samples, and few of the radio galaxies, have contamination 
in the far IR from the radio synchrotron core (Lawrence et al, in prep). (Also
see Hoekstra, Barthel and Hes 1997 and Hes, Barthel and Fosbury 1996 for
additional detailed arguments.)}  
Recently however Gopal Krishna et al 1996 have very cleverly argued that the data are
actually as expected. They assume a torus opening angle increasing with luminosity,
and some properties of radio source evolution taken from independent 
radio astronomy folklore. 
 
  I do not have a strong opinion on which explanation of the 3CR statistics is correct,
but I do understand that this is of great importance physically.  A radio galaxy 
without the ostensibly thermal big blue bump and broad line region in the optical/UV 
would constitute a ``nonthermal AGN." This was the popular explanation for radio galaxies
before the Unified Model removed the compelling observational justification (e.g. 
Begelman, Blandford and Rees 1984).  Some years ago I observed many of these objects
with the Lick spectropolarimeter, with little success in finding hidden quasars.  Of course
a suitable mirror need not be present in all cases.  Thus I am convinced that the robust
test is a search for the waste heat inevitably accompanying this configuration.  Hopefully
the several ISO programs on the 3CR will shed light on this.

  The place to look for the waste heat is in the mid IR ($10-20\mu$).  Starburst emission
isn't strong in this region; emission from nuclear dust tori {\it is} strong there, and also
penetrating.  It is viable to do this study at Keck and we have started taking such data.
The sensitivity is lower than that of ISO but the diffraction-limited images provide
additional information. Also we can select targets to give maximum leverage on this 
problem.  Here I show an observation from our first night, Aug 8 1996, taken with the
``LWS" camera in collaboration with R. Puetter and B. Jones.  To our great surprise
we see {\it two} point sources;  since the only obvious type of powerful compact
$12\mu$ source 
is a dusty torus, there may be two of them!  Of course further observations will be needed
to test the thermal interpretation.

\section {Long Slit Spectropolarimetry at Keck:  Radio Galaxies, BALs and a
Red Quasar}

  I participated in this project but it was initiated by Wil van Breugel and the real work
was done mostly by Andrea Cimatti, Arjun Dey, and Mike Brotherton.  Our goals were to test 
and elucidate the
Unified Model for the aligned radio galaxies at high redshift, and to learn something about
their stellar populations and surrounding gaseous environments.  We also looked at some
related objects as noted in the section title.

  First we observed some galaxies at $Z\sim 1$ which were known to be polarized;  then, motivated
in part by the desire to see the polarization behavior at shorter rest wavelengths, we 
moved on to some $Z \sim 4$ objects without previous polarization data.  Finally we examined a
couple of BALs from the FIRST radio survey, and also a ``Red Quasar."

\subsection  {3C324 at $\bf Z=1.2$, described in Cimatti et al 1996}

  This aligned radio galaxy is $\sim 5 {\tt "}$ long, with lots of structure in the rest UV HST images,
but a relaxed morphology in the rest optical (Longair, Best and Rottgering 1995;
Dickenson, Dey and Spinrad 1995).  We found a power
law spectrum with index $\sim -2.0$ in the rest range $\sim 1600\AA-4200\AA$, a wavelength-independent
11\% polarization perpendicular to the optical/radio axis, and hence a polarized flux
spectral index also of -2.0.  The scattered light extends over several arcsec.  No stellar 
features were found but the data do not extend
to the wavelengths of the strong UV lines.  

  The wavelength independence of P and the (admittedly steep) power
law spectrum of PxF suggest electrons as scattering particles, but a large HII mass would
be required, and the (pretty solid) detection of broad MgII in PxF constrains the temperature
to be $<~10^{6}$K.  Dust scattering is a much more efficient scatterer per gram
of total mass in the Milky Way, but its percent
polarization of the scattered light is generally lower than for electrons.  More 
importantly, it is difficult
to produce a constant polarization and a power law in polarized flux without invoking
somewhat fine-tuned comingled reddening.  See Wills and Hines 1997; Hines and
Schmidt 1997 for data on many dust-scattered and reddened objects.  The required fine-tuning
would also have to account for the scattering phase function, which is highly wavelength-
dependent in the UV.  (See models of e.g. Manzini and di Serego Alighieri 1996; Kartje 1995).
On the other hand a wavelength-independent cross section could be produced intrinsically
perhaps by large grains; however this reduces the cross section per unit mass.  
The dust fraction
may also be lower tens of kpc from a galaxy at high redshift, compared with the local ISM.
As Sandage and Visvanathan once said about M82, our observations have made the object 
a mystery to us.

\subsection  {3C265 at $\bf Z=1.8$, described by Dey et al 1996}

  This case is somewhat similar to 3C324:  a $\sim 5 {\tt "}$ rest UV extent, 11\% polarization independent
of wavelength, and power-law total and polarized flux, but here with spectral index -1.1.  
(The total flux also shows a very broad absorption line shortward of CIV and 
quite detached from it!)  Assuming this is Thomsom-scattered light from 
a hidden quasar which is like
the similar lobe-dominant ones at this
redshift in the 3CR catalog (i.e. the Unified Model scenario), we
estimated $\sim 2\times 10^{11}$Mo in HII would be required in the scattering
``cone" alone.
The SNR is inadequate to
check for polarized broad emission lines, so the optical data do not preclude very hot
gas;  however such gas would overpredict the Rosat flux (see Dey et al 1996).  Figure 9 in our
paper shows the comparison of the total flux spectrum from $1500\AA$ to
$3100\AA$ to that of NGC1068,
considered to be electron scattered, and they are very similar, although it's possible 
that the FeII contribution is different and so we are being misled.  Again all evidence points
to electron scattering except plausibility - the great mass of warm gas couldn't be very
long lived.  It would also overpredict the recombination emission unless it is distributed
very smoothly.  Luckily P. Eisenhart et al have HST imaging polarimetry pending for this one.

\subsection {3C13 at $\bf Z=1.4$ and 3C356 at $\bf Z=1.1$, described by Cimatti et al 1997}

  3C13 is faint but we can at least say it has $\sim 8\%$ polarization at the expected position
angle.  3C356 is brighter, and is known for showing a strong alignment effect even in the
rest optical (Eales and Rawlings 1990;  Lacey and Rawlings 1994).   
The bright northern component has P increasing with frequency to
$\sim 15\%$, again a power-law polarized flux distribution though this time with index -0.2,
and a polarized MgII broad emission line.  The Rosat data imply a massive cooling flow
with the right electron scattering optical depth according to Crawford and
Fabian 1996), but any
very hot gas cannot contribute to the polarized MgII line.  

\subsection {4C41.17 at $\bf Z=3.8$, described by Dey et al 1998, and 7C1909+722 at
$\bf Z=3.6$
(paper in prep)}
 
  At this point we were more interested in other aspects of the project than in the
implications for Unification:  Are there really massive clouds of warm (not hot) electrons?
What about starlight?  Thus we observed a couple at higher redshift, where any stellar lines 
should be apparent, and where dust polarization and polarized flux should show a very 
strong wavelength dependence.
The first target, 4C41.17, turned out to be unpolarized and to have strong stellar
absorption lines.  A spectacular HST images was shown at the meeting by van Breugel.
Clearly in this case the alignment effect is due to starlight, and the
star formation is truly prodigious, easily producing the stellar mass of a luminous
galaxy in a dynamical time.  There may be some scattered quasar light in the observed
red region, where there is a possible broad CIII component and only poor polarization
limits.  If so the scattering doesn't rise rapidly at the shortest wavelengths, as
expected for Galactic dust.  The 7C galaxy in the section title was discovered by van Breugel
and Hurt based on its steep radio spectrum, and is qualitatively similar to 4C41.17.

  In a further attempt to test dust vs electrons as the scatterers, 
we applied unsuccessfully to get NICMOS time for some polarization images; however ground-based IR data,
combined with optical data, can be constraining (Knopp and Chambers 1997).

\subsection  {3C68.1, a red quasar at $\bf Z=1.2$, described in Brotherton et al 1998}

  The ``red quasars" have historically been puzzling, and seem to be a mix of objects with
strong synchrotron components extending into the near IR, and highly reddened objects often
with accompanying dust scattering.  This one has a spectral index of -6 in the observed
optical, and a lobe-dominant radio source with a weak synchrotron core.  Indeed the optical
light turns out to be a combination of a highly reddened direct view of a quasar with
a moderately-reddened dust-scattered view of the same object.  It is similar to the
radio quiet object IRAS 13349+2438 (see Wills et al 1992).

\subsection  {BAL quasars from the FIRST radio survey, 0840+3633 and 1556+3517,
described in Brotherton
et al 1997}

  These are borderline radio loud BALs, discovered by Becker's group based on the deep VLA
radio survey called FIRST.  Interestingly, they are both low ionization, with strong FeII
absorption.  Past work, including extensive spectropolarimetry, has lead to a picture
of BALs as nearly edge-on, borderline objects in the Unified Model, with no definite
explanation for the lack of BALs among classical radio-loud objects.  0840+3633 fits
the pattern to some extent previous observations:  the polarization rises to 4\% at
$2000\AA$, 
possibly diluted at longer wavelengths by the small blue bump.  The troughs are much more
highly polarized, as usual, suggesting that the scattered light finds a path
which partially
avoids the broad absorption line clouds.  However the magnitude and position angle
withing the FeII* (excited) troughs shows a lot of complicated behavior, with some more
polarized and some less polarized than the continuum!

  The properties of 1556+3517 are really new:  the continuum polarization rises with
decreasing wavelength to 13\% at $2000\AA$.  Unlike the other BALs it shows BLR polarizations
consistent with that of the surrounding continuum, and {\it no} polarization in the FeII
troughs!  Perhaps all the nuclear and BLR light passes through the BAL clouds, and the
unpolarized light in the troughs is starlight from the host galaxy, but this is just 
a guess.

  Are BALs really nearly edge-on?  One promising approach to this question has only
lead to more difficulty.  Radio spectral slopes are now available for many radio quiet
AGN (Antonucci and Barvainis 1988, 1989; Barvainis, Lonsdale and Antonucci
1996)), and they are a mixture of types as for radio loud AGN.
It has been argued recently that the flat ones are beamed versions of the steep ones
(Falcke, Sherwood and Patnaik 1996).  But the BALs are a rather even mixture of the two 
types, (Barvainis and Lonsdale 1997), 
showing that they are fairly isotropic, or else that we are really confused about 
the causes of the spectral slopes in radio quiet AGN.  Perhaps with great difficulty
radio axes can be obtained for some BALs; this would help in interpreting the polarization.

\section {Nature of the Far-IR Based on Radio Quiet Quasars}

  Many detections of quasars and high redshift radio galaxies are being reported 
in the millimeter region
of the spectrum.  I'd like to review briefly what we know about the similar sources
among radio quiet quasars and IRAS galaxies, which are analogous to these
two classes.  The subject is further along for the radio quiets.  

  First, of course,
the field was galvanized by IRAS 10214+4724 at $Z=2.3$, detected in the far IR and then 
in the millimeter continuum and various millimeter lines.  Lensing may have been crucial for boosting 
the flux to observable levels (but see Scoville et al 1995).
This famous object needs a name and I call it the Mother of All IRAS Galaxies, referring to
a contemporaneous news event.  Much of the line emission, and hence probably the dust,
is very compact, though some is apparently resolved on $\sim 10$kpc scales 
(Brown and vanden Bout 1991; Solomon et al 1992a,b; Scoville et al 1995).
High optical polarization was discovered by Lawrence et al 1992, and subsequent Keck spectropolarimetry
showed that it is an obscured quasar (Goodrich et al 1992).  In fact what we observe in the optical is
mainly the lensed image of the scattering mirror.  Two more ``hyperluminous" IRAS    
galaxies have been reported and both are also hidden quasars revealed 
in the polarized flux spectra.

  In 1992 we reported detection of the Cloverleaf ($Z=2.6$, quad lensed, BAL)
quasar at $350\mu$,
$450\mu$, and $800\mu$ in the observed frame (Barvainis, Antonucci and Coleman 1992).  
This was the first such observation that I know
of, though Chini et al had already shown that many quasars must have steep spectra between
the $100\mu$ IRAS point and 1.3mm.  The observations were the culmination of a five
year struggle with the old Ukt14 bolometer at the JCMT;  they would be trivial now
with the Scuba bolometer array.  Perhaps we shouldn't have bothered with Ukt14;  
in the immortal
words of my former thesis advisor, ``Why walk when the bus is coming?"  

  Our original goal was to see whether quasars follow the far IR/CO luminosity
correlation as IRAS galaxies do, suggesting that the quasar far IR source is
thermal, and testing the proposition that radio quiet quasars and IRAS galaxies
differ only in orientation.  They do follow that relation quite well (e.g. Alloin
et al 1992).  (Also the Cloverleaf and MOAIG have the same SED in the IR range,
just diverging in the optical as expected (Barvainis et al 1995).  Whenever 
one shows the far IR vs CO luminosity diagram, there is an opportunity to show
also the one by Kennicutt (1990).  He shows data for ordinary galaxies but extends
the luminosity range by including a burning cigar, a Jeep Cherokee, the 1988
Yellowstone forest fire, Venus and the universe within the horizon.  The
correlation is
very tight and convincing and there are no unplotted upper limits so it is
definitely real.  

  Followup observations detected several transitions of CO, HCN and CI, including
detailed high-resolution mapping of CO 7-6 which resolves the four
images  (Yun et al 1997; Alloin et al 1997; Kneib et al 1997).  Modeling by Phil Maloney shows that the
H2 mass is considerably less than indicated by the conversion factor for Galactic
molecular clouds, as a result of high temperature and moderate optical depth.
Phil has also inferred several times more mass in an HI region based in part on 
the CI lines.    The CO source is
only $\sim 600$ pc in radius, where we use the high resolution as well as the high flux
afforded by the gravitational telescope.  The data suggest a rotating disk, with 
dynamical mass of $\gtwid 10^{10}$ Lo, consistent with that of the CO region.  One interesting 
loose end however is
that we have detected a large spatial extent for some of the 1.3mm continuum flux,
visible with ``5-point" observations with the 30-m IRAM Pico Vileta dish.  It remains
to be determined what that is all about.

  There is an interesting literature on the CO flux to H2 mass conversion factor.
Various lines of reasoning lead to the conclusion that this conversion factor is
low in IRAS galaxies (and quasars), as discussed in detail by Maloney 1990 
and Solomon et al 1997 for example.
A seeming proof was given by Shier, Rieke and Rieke 1994 for luminous IRAS galaxies:  
the conventionally
deduced H2 masses exceed the estimated dynamical masses in that case.  However it's not certain
that the near-IR wavelengths used here were sufficiently penetrating.
Moreover Scoville et al (1995)
interpreted the same situation in the MOAIG differently:  they proposed
that the mass is real, but that the gas is supported partially by radiation 
pressure rather than centrifugal force!  Detailed modeling using many transitions
considerably reduces the uncertainties however and as noted the models for the
Cloverleaf indicate a much lower than standard conversion factor for that object
(Barvainis et al 1997).

  Many more quasars, high-redshift IRAS galaxies, and radio galaxies are being
detected in the mm and in the CO lines: e.g. the lens MG0414+0534 by our group
(Barvainis et al 1998);  53w002 (Scoville et al 1998); BRI 1335-0415 at $Z=4.4$ 
(Guilloteau et al 1997 preprint);
and the $Z=4.7$ quasar
BR1202-0725 {\it and an otherwise invisible companion} (Ohta, Omont).  It has been pointed out
that the K-corrections in both continuum (Blain and Longair 1993) and rotational lines
(Solomon, Radford and Downes 1992)
are so sensitive to $Z$ that more distant objects can be considerably brighter
than less distant ones.  (See also Barvainis' 1996
News and Views article.)
Thus the big new dishes and arrays will open up the
early universe to the study of a component which is an important part of the
baryonic mass of forming galaxies.

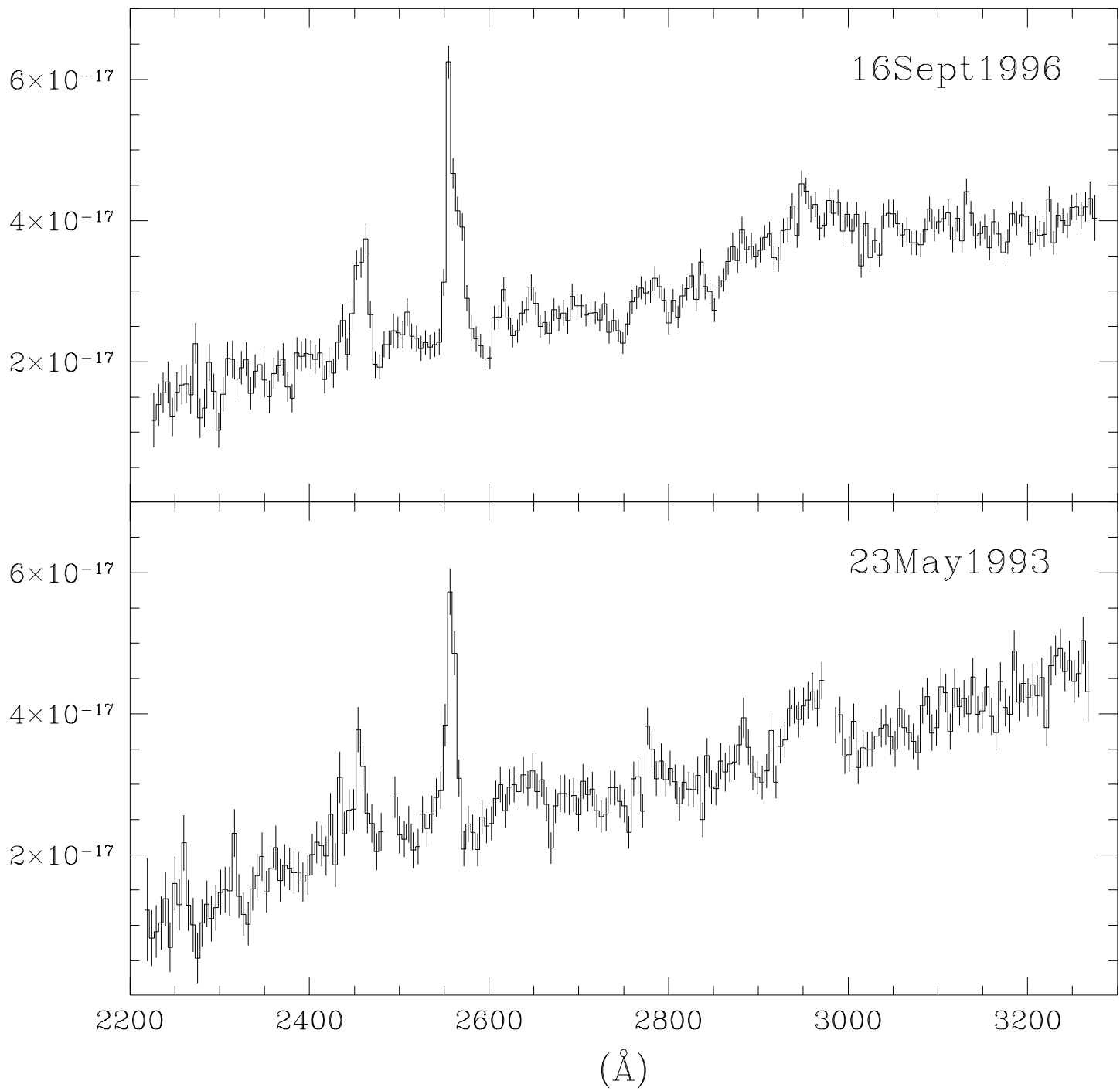
\end{document}